\documentclass[preprint]{aastex}
\newfont{\smcap}{cmcsc10 at 12pt}
\newcommand{\au}{\mbox{\smcap au}}
\newcommand{\cm}{{\rm cm}}

\newcommand{\hg}{H_{\rm g}}
\newcommand{\hp}{H_{\rm p}}
\newcommand{\Kelvin}{~{\rm K}}
\newcommand{\LE}{L_{\rm E}}
\newcommand{\lin}{{\cal L}}
\newcommand{\nuT}{\nu_{\rm T}}
\newcommand{\pd}{\partial}
\newcommand{\Rey}{{\rm Re}_*}

\newcommand{\rp}{r_{\rm p}}
\newcommand{\rhog}{\rho_{\rm g}}
\newcommand{\rhop}{\rho_{\rm p}}
\newcommand{\rhos}{\rho_{\rm s}}
\newcommand{\s}{{\rm s}}
\newcommand{\Sc}{{\rm Sc}}
\newcommand{\Sigmag}{\Sigma_{\rm g}}
\newcommand{\Sigmap}{\Sigma_{\rm p}}
\newcommand{\ts}{t_{\rm s}}
\newcommand{\vy}{\mbox{\boldmath $y$}}

\newcommand{\vk}{V_{\rm K}}
\newcommand{\yr}{{\rm yr}}
\renewcommand{\apj}{{\it Astrophys. J.}}
\newcommand{\icarus}{{\it Icarus}}
\renewcommand{\mnras}{{\it Mon. Not. R. Astron. Soc.}}
\newcommand{\rmp}{{\it Rev. Mod. Phys.}}

\shortauthors{GOODMAN AND PINDOR}
\shorttitle{DUST-LAYER INSTABILITY}

\begin{document}
\title{Secular Instability and Planetesimal Formation in the Dust Layer}

\author{\smcap J. Goodman \\ B. Pindor}
\affil{Princeton University Observatory, Princeton, NJ 08544}
\email{{\rm Email:} jeremy@astro.princeton.edu}

\begin{abstract}
Late in the gaseous phase of a protostellar disk, centimeter-sized
bodies probably settle into a thin ``dust layer'' at the midplane. A
velocity difference between the dust layer and the gas gives rise to
turbulence, which prevents further settling and direct gravitational
instability of the layer.  The associated drag on the surface of the
layer causes orbital decay in a few thousand years---as opposed to a
few hundred years for an isolated meter-sized body.  Within this
widely-accepted theoretical framework, we show that the turbulent drag
causes radial instabilities even if the selfgravity of the layer is
negligible.  We formulate axisymmetric, height-integrated dynamical
equations for the layer that incorporate turbulent diffusion of mass
and momentum in radius and height, vertical settling, selfgravity, and
resistance to compression due to gas entrained within the dust layer.
In steady-state, the equations describe the inward radial drift of a
uniform dust layer.  In perturbation, overdense rings form on an
orbital timescale with widths comparable to the dust-layer thickness.
Selfgravity is almost irrelevant to the linear growth rate but will
eventually fragment and collapse the rings into planetesimals larger
than a kilometer.  We estimate that the drag instability is most
efficient at 1 AU when most of the ``dust'' mass lies in the size
range 0.1-10 meters.
\end{abstract}

\section{INTRODUCTION}

Meteoritic evidence and theoretical considerations indicate that
planet formation begins with collisional agglomeration of dust
particles into larger bodies during the lifetime of a gaseous
protostellar disk ($\lesssim 10^7~\yr$).

Because the gas is partly supported against the gravity of the central star
by pressure, the orbital velocity of the disk 
is slightly lower than the keplerian circular velocity ($\vk=\sqrt{GM/r}$).
If $c_s$ is the sound speed of the gas, then
\begin{equation}\label{etadef}
\eta\equiv \frac{\vk-V_{\rm gas}}{\vk}\approx
 -\frac{r}{2\rhog\vk^2}\frac{\partial P}{\partial r}
 \sim\left(\frac{c}{\vk}\right)^2.
\end{equation}
Solid bodies bodies couple to the gas by aerodynamic drag and orbit
at intermediate velocities.
Small grains move almost exactly with the gas, while
large planetesimals move nearly at $\vk$.
In both limits, drag causes very little dissipation of
orbital energy.  Dissipation is maximal for bodies
of intermediate size.
At 1~AU in a minimum-mass solar nebula \citep{HNN},
the orbital decay time is only a few hundred years for
particle radii $\rp\sim 1$~m (cf. \S 2).  Therefore,
the processes responsible for growth must move bodies
rapidly through the intermediate size range.  Since the orbital decay
time scales roughly linearly with $\rp$ above $\sim 1~\mbox{m}$,
planetesimals larger than
$\rp\sim 10~\mbox{km}$
experience little decay during the lifetime of the nebula.

While collisional agglomeration may be sufficient to reach
$\gtrsim~\mbox{km}$ size before the orbits decay, collective instability
is an attractive alternative.
\cite{Safronov} and \citet[henceforth GW]{GW}
suggested that planetesimals
may form by gravitational instability.
As an initially turbulent disk becomes quiescent, or as the
solid bodies grow to radii $\rp\sim 1~\cm$,
the particles settle into an increasingly dense
layer---the ``dust layer''---at the midplane.
When the volume-averaged density of the dust layer
approaches $\rho_*\equiv 3M/4\pi a^3$,
where $M$ is the mass of the star and $a$ is the orbital radius,
the layer becomes gravitationally unstable.
Self-gravity then fragments the layer into planetesimals with
radii $\sim 1~\mbox{km}$.

Gravitational instability along the above lines is now regarded
as unlikely. The dust layer would have to be very
much thinner than the gas disk, since the mass fraction of solids
at solar abundances is small ($\sim 10^{-2}$), and since
the gas density itself is already $\sim 10^{-2}\rho_*$
in a minimum-mass solar nebula.
On the other hand,
the velocity shear between the dust layer
and the gas probably causes turbulence, even if there is no
other source of turbulence in the gas disk.
Although GW called attention to this
``boundary-layer'' turbulence, they considered its implications
only for the exchange of momentum between the dust layer and the gas.
\cite{Weid80} pointed out that turbulence will also
tend to limit the density of the dust layer by mixing.  More recently,
\citet[henceforth CDC]{Cuzzi} have confirmed this with a detailed
analysis particle settling, shear-driven turbulence.  and turbulent
vertical mixing.  These authors conclude that gravitational
instability is unlikely to occur at $1~\au$ until $\rp\gg 1~\mbox{m}$,
if at all.

Even if direct gravitational instability does not occur, there may be
secular instabilities---that is, instabilities relying upon
dissipation.  GW mentioned the possibility of secular effects but did
not elaborate.  They may have had in mind the following axisymmetric
instability: Consider a steady background state in which the surface
mass density of the dust layer ($\Sigmap$) is uniform on scales $\ll
r$ and drifts radially inward at a rate proportional to the drag
against the gas.  Let there be a positive perturbation in $\Sigmap$ on
a ring of radial width ($\Delta r$) smaller than the thickness of the
gas layer ($\hg$) but larger than that of the dust layer ($\hp$).  The
(slight) selfgravity of the ring reinforces the gravitational
attraction of the star on the dust layer at the outer edge of the
ring, and has the opposite effect at the inner edge of the ring.
These gravitational perturbations must be balanced by pressure
gradients or centrifugal forces.  Since $\hp<\Delta r<\hg$, the
orbital speed of the gas is much less affected than the orbital speed
of the dust, which increases at the outer edge of the ring and
decreases at the inner edge.  The outer (inner) edge suffers greater
(less) drag and a larger (smaller) inward radial drift rate.  This
reinforces the perturbation in $\Sigmap$.
Drag is clearly essential: without it,
contraction of the dense ring would be prevented for $\Delta r\gg\hp$
by the difference in specific angular momentum between the inner and
outer edges of the ring; but drag changes the angular momentum of the dust.
The planform of the secular instability is expected to be stationary
in a frame drifting inward with the dust layer.
\cite{Saf91} remarked that the layer will probably be 
stabilized by the difference in radial drift speed among bodies of
different size.  This assumes that the drag acts separately on
individual bodies.  Many theoretical studies of the dust layer
(including GW and CDC), however, assume that the drag
acts collectively because the gas
\emph{within} the layer is entrained by the greater mass of solids.
In \S4, we shall find that whether the particles feel the drag
individually or collectively depends upon their size.
For the present, we take the drag to be collective.

The original intent of the present work was to study the secular
gravitational instability described above.
We expected to find a slowly growing mode (compared to the orbital
frequency) on scales $\Delta r\gg\hp$, whose growth rate would 
tend to zero with $\rho/\rho_*$.
But our analysis uncovered a rapid, small-scale, drag-driven instability 
that does not require any selfgravity, although it does require that
the drag be collective.

The structure of this paper is as follows.  In \S2, we present
height-integrated nonlinear equations for the axisymmetric evolution
of the dust layer based largely upon the work of CDC,
and we cast these equations into dimensionless form.  We also discuss
our assumptions for the nebular parameters, and we find steady
solutions to the equations in which the surface density is locally
uniform.  The linear stability of the steady solutions is examined in
\S3, first analytically and then semi-numerically.  The most rapid
instabilities are driven by drag rather than selfgravity.  The
following section, \S 4, explores a number of important issues,
including the collective nature of the drag force, the relative
importance of the instability for building planetesimals compared to
straightforward collisional agglomeration, and the expected
planetesimal mass if the instability dominates.
\S4 also briefly compares this work with some notable
previous investigations of the stability
of protoplanetary disks that treat gas and dust as dynamically distinct
components \citep{CFM,SN88,NVC}. \S5 sums up.

\section{PHYSICAL MODEL AND GOVERNING EQUATIONS}
\label{sec:eqns}

The purpose of this section is to develop model equations for
axisymmetric radial perturbations of the dust layer.
To avoid having to treat two spatial dimensions at once, we
adopt a height-integrated approximation based upon the results
of CDC, whose notation we generally follow.
Important physical influences on the evolution of the dust layer
include turbulent drag against the gas; turbulent mixing,
horizontally as well as vertically; gas pressure (which resists
compression of the layer); and selfgravity.


We are concerned with a late phase in the evolution of the
gaseous protoplanetary nebula when accretion and its attendant
turbulence have subsided, allowing the dust to settle
into a thin layer of half-thickness $\hp$ very much less
than that of the gas, $\hg$ \citep{Safronov,GW}.
(In this paper, ``dust'' refers indiscriminantly to all
solids smaller than a few meters.)
Within the dust layer, the volume-averaged mass density 
of the dust ($\rhop$) exceeds that of the gas ($\rhog$), as
shown below, but the column density of the gas
($\Sigmag\approx 2\hg\rhog$) greatly exceeds that of the dust
($\Sigmap\approx 2\hp\rhop$).  At solar abundance, 
$\Sigmag/\Sigmap\approx 200$ if the dust contains all materials
condensible at temperatures $\sim 300\Kelvin$.

We assume a standard minimum-mass
solar nebula with the following radial scalings:
\begin{eqnarray}\label{MMSN}
T(r) &=& 280~{\rm K} \left(\frac{r}{\au}\right)^{-0.5}\nonumber\\
\Sigmag(r) &=& 1700~{\rm g~cm}^{-2}\left(\frac{r}{\au}\right)^{-1.5}
\nonumber\\
\rhog(r) &=& 1.7\times 10^{-9}~{\rm g~cm}^{-3}
\left(\frac{r}{\au}\right)^{-2.75}.
\end{eqnarray}
For ease of reference, we summarize the more important physical
parameters in Table I.
\marginpar{\small\it Table I here}

\subsection{Turbulent drag on the dust layer}

In steady state, the dust layer orbits nearly at the
the keplerian velocity $\vk = \sqrt{GM/r}$, where
$M$ is the mass of the protostar, and $r$ is the orbital
radius.  (Unless otherwise stated, $M=1~M_\odot$
and $r= 1\au$.)  Because of its radial pressure gradients,
the gas orbits at the slightly lower velocity $(1-\eta)\vk$
determined by equation (\ref{etadef}).
For nominal parameters, $\eta\approx 10^{-3}$ [Table I].

Because of the orbital velocity difference above,
GW argued that the interface between
the dust layer and the relatively pristine gas should be turbulent,
even if the nebula is otherwise quiescent,
by analogy with the standard laboratory situation of a thin solid
plate placed edgewise in a nearly inviscid, incompressible flow.
The drag on such a plate due to turbulent momentum transport is
\begin{equation}\label{Fdrag}
\vec F_{\rm drag}=-\frac{\rhog|\Delta\vec v|}{\Rey}\Delta\vec v,
\end{equation}
where $\Delta v$ is the velocity difference between the plate and
the undisturbed flow, and the ``Reynolds number''
$\Rey$ is an empirical dimensionless constant.
The drag law is usually stated in scalar form,
$F_{\rm drag}=-\rhog(\Delta v)^2/\Rey$; the vectorial form above
assumes that the force is antiparallel to the relative velocity.
Citing laboratory measurements, GW took
$\Rey\approx 500$, but CDC advocated
$\Rey\approx 45-180$.  We take $\Rey=180$ as a reference value
but are careful to note how things scale with this number.

The turbulent transfer of momentum from the dust layer (or flat plate)
to the surrounding medium is effectively diffusive and can
be described by a \emph{turbulent viscosity} $\nuT$, with dimensions
of $(\mbox{length})^2(\mbox{time})^{-1}$.
In order to reproduce the drag (\ref{Fdrag}), one needs
$\nuT\approx L\Delta v/\Rey$, where $L$ is the vertical thickness of the 
\emph{turbulent boundary layer}---the layer within which the velocity,
though fluctuating, is closer to that of the dust
layer than to that of the undisturbed gas.  As noted by
GW, $L$ should be comparable to the Ekman length
constructed upon $\nuT$ and the orbital angular velocity $\Omega$:
\begin{equation}
\LE\ =\left(\frac{\nuT}{\Omega}\right)^{1/2}.
\end{equation}
Viewed in inertial coordinates, the direction of the
stress exerted by the dust layer on the gas rotates during the orbit,
and $\LE$ is the vertical distance over which horizontal momentum diffuses
in time $\Omega^{-1}$.  Similar considerations govern the thickness
of the boundary layer at the base of the Earth's atmosphere \citep{Pedlosky}.
As a generic measure of the boundary-layer thickness,
CDC take $L=c_L\LE$ with $c_L=1.5$.
Since $\nuT\approx L\Delta v/\Rey$, it follows that
\begin{equation}\label{nuT}
\nuT\approx\frac{c_L^2\Delta v^2}{\Omega\Rey^2},
\end{equation}
and also that
\begin{equation}\label{L}
L\approx \frac{c_L\Delta v}{\Rey\Omega}.
\end{equation}

The factors involving $\Rey$ in eqs.~(\ref{Fdrag}), (\ref{nuT}), and
(\ref{L}) represent significant corrections to naive dimensional
analysis.  They are semi-empirical, based upon terrestrial
experience that may not be a reliable guide to the situation we wish
to study. However, it seems difficult to improve upon these factors
without resort to an ambitious three-dimensional hydrodynamic simulation.

\subsection{Thickness of the dust layer}

The equilibrium half-thickness ${H_{\rm p,0}}$
of the dust layer results from competition between
turbulent stirring and gravitational settling.
Following CDC, we assume that $\hp$ is comparable to
the \emph{1\% boundary-layer thickness}---the height above the midplane
at which the mean azimuthal velocity is within $1\%$ of its asymptotic
value far above the dust.
Also following CDC, we take this thickness to be larger
than $L$ by a factor $c_\delta\approx 2.5$.  Thus,
\begin{equation}\label{hpest}
{H_{\rm p,0}}= c_\delta c_L \frac{\Delta v}{\Omega\Rey}.
\end{equation}
Unless otherwise noted, we adopt eq.~(\ref{hpest}) for the dust-layer 
thickness.  We believe that it is appropriate for dust-particle
sizes of order $10~\cm$.
But the layer is likely to be somewhat thicker (thinner)
if the particles are much smaller (larger) than this.

When all particles are small they will be tightly
coupled to the gas within the dust layer by drag (see below).
In this limit, where the settling time is effectively infinite and
the slightest turbulence is sufficient to loft the dust,
the equilibrium vertical distribution of dust is likely to occur
at the point of marginal shear instability, absent other sources of
turbulence.
The following is a simplified
estimate of the thickness under these assumptions; a more careful
treatment has been given by \cite{Sekiya}. 
Shear instability is expected only if the Richardson number
\begin{equation}\label{Richardson}
Ri \equiv \frac{(\pd\Phi/\pd z)(\pd\ln\rho/\pd z)}{(\pd v/\pd z)^2}
\le \frac{1}{4}.
\end{equation}
The vertical gravitational potential $\Phi(z)$ is $\Omega^2 z^2/2$, 
and the horizontal velocity $v(z)$ (relative to the keplerian value $\vk$)
is determined by the geostrophic balance
\begin{equation}
2\rho\Omega v = \frac{\pd P}{\pd r}.
\end{equation}
For simplicity, we take gaussian profiles for both the dust and
the gas, $\rhop(z)=\rhop(0)e^{-z^2/2\hp^2}$ and 
$\rhog(z)=\rhog(0)e^{-z^2/2\hg^2}$, with $\hp\ll\hg$.
Then since $v(z)\to -\eta\vk$ when $z\gg\hp$,
\begin{eqnarray*}
\rho(z) &=&\rhog(z)+\rhop(z)\approx\rho(0)\left(1-f+f e^{-z^2/2\hp^2}\right)\\
v(z)&\approx& -\eta\vk\left(1 + \frac{f}{1-f}\,e^{-z^2/2\hp^2}\right)^{-1},
\end{eqnarray*}
where $f\equiv \rhop(0)/\rho(0)$ is the fractional contribution of the
dust to the total density at the midplane.
With these assumptions, $Ri(z)\approx Ri(0)$ 
within one scale height of the midplane, and  increases at higher altitudes.
\begin{displaymath}
Ri(0) = \left(\frac{\Omega\hp}{\eta\vk}\right)^2\,\frac{(1-f)^2}{f^3}
\end{displaymath}
On the other hand,
\begin{displaymath}
\frac{\rhop(0)\hp}{\rhog(0)\hg}=\frac{\Sigmap}{\Sigmag}\equiv{Z_{\rm p}},
\end{displaymath}
where ${Z_{\rm p}}\approx 0.005$ is the fractional abundance of dust.
Together with $\hg=c_s/\Omega\approx\eta^{1/2}\vk$, these relations
imply that the condition $Ri(0)\le 1/4$ is equivalent to
\begin{displaymath}
\frac{\rhop(0)}{\rho(0)}\ge\left(\frac{4{Z_{\rm p}}^2}{\eta}\right)^{1/3}
\approx 0.46,
\end{displaymath}
\emph{i.e.} about equal contributions to the
midplane density from dust and gas.  It follows that
the dust-layer thickness in the limit of very small particles is
\begin{equation}\label{margturb}
(\hp)' = \frac{1-f}{f}\,{Z_{\rm p}}\hg~\lesssim~ 0.18\frac{\eta\vk}{\Omega}
\end{equation}
This is about $9$ times larger than our standard estimate (\ref{hpest})
if we take $\Rey=180$ and $\Delta v=\eta\vk$.
In the small-particle limit where (\ref{margturb}) holds, there may be very
little drag on the dust layer, since the turbulence is only incipient.

Larger particles are imperfectly coupled
to the gas, and significant turbulence will be required to keep
them from settling.
A dimensionless measure of the strength of drag coupling
is $\Omega\ts$, where the  ``stopping time'' $\ts$ is the decay time 
of the velocity of an isolated  particle relative to the gas if
subject to no forces other than drag.
As discussed by CDC and others, there are at least
three distinct drag regimes of interest, distinguished by the
relative sizes of the the particle radius ($\rp$), the molecular
mean-free path ($\lambda$), the sound speed ($c$), and the
relative velocity between particle and gas ($v_{\rm rel}$).
Particles with radii $\rp\le9\lambda/4$ are in the ``Epstein''
regime, where the velocities of gas molecules striking the
particle are uncorrelated with that of the particle itself,
and the drag force is $4\pi\rp^2\rhog c v_{\rm rel}/3$.
If $\rp>9\lambda/4$ but $\rp<\lambda c/(4 v_{\rm rel})$, there is
a laminar viscous flow over the particle, and the drag
is given by Stokes' formula $6\pi\mu\rp v_{\rm rel}$, where
$\mu\approx\rhog\lambda c/2$ is the dynamical viscosity of the gas.
At still larger values of $\rp$ and/or $v_{\rm rel}$, the flow
over the particle is turbulent and the drag force increases
faster than linearly with $v_{\rm rel}$, eventually quadratically.
The mean-free path $\lambda\approx 1 (r/\au)^{2.75}\cm$ at the midplane
of the nebula (\ref{MMSN}).
Thus
\begin{equation}\label{tstop}
\ts\approx\cases{1.2\times10^4(\rp/\cm)(r/\au)^3~\s & ``Epstein'',\cr
            5.3\times10^3(\rp/\cm)^2(r/\au)^{2.75}~\s &``Stokes'',\cr}
\end{equation}
with the dividing line between the two regimes at
$\rp\approx 2.3(r/\au)^{2.75}\cm$.  Presuming that
$v_{\rm rel}<\eta\Delta v$, the turbulent regime requires
$\rp> 10(r/\au)^{2.5}\cm$.

Since the vertical gravitational acceleration 
is $\Omega^2 z$, a particle settling through quiescent gas reaches a terminal
velocity $\Omega^2 z/\ts$ provided $\Omega\ts<1$, so that $z$
decays exponentially at the rate 
\begin{equation}\label{gamma_s}
\gamma_{\rm s}^{-1}\equiv (\Omega^2 t_{\rm s})^{-1}.
\end{equation}
Turbulence will resist this settling.
Following CDC, we describe turbulent stirring
by a mass-diffusion coefficient 
\begin{equation}\label{numass}
\nu_{\rm mass}=\frac{\nuT}{\Sc}.
\end{equation}
where the ``Schmidt number'' $\Sc\ge 1$ describes the efficiency
with which turbulent eddies are able to pick up the particles;
$\Sc$ increases with particle size and with orbital radius
in the nebula [cf. Fig. 2 in CDC].
Therefore, still another estimate
of the dust-layer thickness, $\hp''$,
follows from equating the diffusion rate $\nu_{\rm mass}/\hp^2$
to the settling rate $\Omega^2\ts$:
\begin{equation}\label{hpp}
\hp''=\left(\frac{\nuT}{\Sc\Omega^2\ts}\right)^{1/2}=
\frac{\hp}{c_\delta\sqrt{\Sc\,\Omega\ts}}.
\end{equation}
For 10-cm particles at $1~\au$, $\Sc\approx 3$ and
the thicknesses (\ref{hpest}) and (\ref{hpp}) are comparable.
For much smaller particles, where $\hp''>\hp$,
the turbulence and its associated drag are probably incompletely developed,
so that the estimate (\ref{margturb}) is probably more accurate.
For larger particles or larger orbital radii, $\hp''$ is smaller than $\hp$,
so that in using eq.~(\ref{hpest}) for the thickness, we are probably
underestimating the effects of selfgravity.

\subsection{Vertically integrated equations of motion}

Here we lay out the equations of motion for the dust layer in
a height-averaged approximation.
The gas above and below the dust layer is taken to be undisturbed,
on the grounds that we are concerned with radial scales $\Delta r$
that are somewhat greater than $\hp$ but $\ll\hg$.  
Disturbances in the pressure and density of overlying gas
are expected to equilibrate on a timescale 
\begin{displaymath}
\frac{\Delta r}{c}\approx \frac{\Delta r}{\hg}\Omega^{-1}\ll\Omega^{-1},
\end{displaymath}
whereas we shall be concerned with motions of the dust layer
over times $\gtrsim\Omega^{-1}$.
We shall allow, however, for the pressure of the gas
trapped within the dust layer.

We write $U$ and $V+V_K$ for the vertically-averaged
radial and azimuthal velocities of the dust layer:
\begin{eqnarray*}
\Sigmap(r,t)&=&\int\limits_\infty^\infty\rhop(r,t,z)dz,\\
\left\{{ U(r,t)\atop V(r,t)}\right\}
&=&\frac{1}{\Sigmap(r,t)}\int\limits_\infty^\infty
\rhop(r,t,z)\left\{{v_r(r,t,z)\atop v_\phi(r,t,z)-\vk(r)}\right\}dz.
\end{eqnarray*}
Continuity of dust mass is expressed by
\begin{displaymath}
\frac{\partial\Sigmap}{\partial t} + \frac{1}{r}\frac{\partial}{\partial r}
\left[r\left(U\Sigmap - \nu\frac{\partial\Sigmap}
{\partial r}\right)\right] = 0.
\end{displaymath}
The term involving $\nu$ reflects our assumption that
if the turbulence mixes the particles vertically, then it also mixes
them radially.
For want of a better hypothesis, we take $\nu=\nu_{\rm mass}$ 
[eq.~(\ref{numass}].
On lengthscales $\Delta r\ll r$,
the metric factors $r^{-1}$ and $r$ surrounding the
radial derivative above can be neglected, so our mass equation becomes
\begin{equation}\label{masseqn}
\frac{\partial\Sigmap}{\partial t} + \frac{\partial}{\partial r}
\left(U\Sigmap - \nu\frac{\partial\Sigmap}
{\partial r}\right) = 0.
\end{equation}

Next, we develop equations for the velocity components $V$ and $U$.
In axisymmetry the only azimuthal force is drag, so that if radial 
diffusion is neglected for the moment,
then the specific angular momentum of the dust evolves as
\begin{displaymath}
\left(\frac{\partial}{\partial t}+U\frac{\partial}{\partial r}\right)
[r(V+\vk)]
= -r\,\frac{\rhog\Delta v}{\Sigmap\Rey}\,(V+\eta\vk),
\end{displaymath}
or equivalently, since $V/r \ll\Omega$,
\begin{equation}\label{eulerV}
\left(\frac{\partial}{\partial t}+U\frac{\partial}{\partial r}\right)
V = -\frac{\Omega U}{2} 
-\,\frac{\rhog\Delta v}{\Sigmap\Rey}\,(V+\eta\vk),
\end{equation}
where 
\begin{equation}\label{Delta_v}
\Delta v \equiv|\Delta\vec v|= \sqrt{U^2+(V+\eta V_K)^2}.
\end{equation}

The dynamical equation for $U$ incorporates coriolis force (the residual
of centrifigal and central forces), drag, self-gravity, and pressure:
\begin{equation}\label{eulerU}
\left(\frac{\partial}{\partial t}+U\frac{\partial}{\partial r}\right)U
= 2\Omega V -\frac{\rhog|\Delta v|}{\Sigmap\Rey} U
-\frac{\partial\psi}{\partial r}
-\frac{1}{\Sigmap}\frac{\partial\Pi}{\partial r}.
\end{equation}
An adequate local approximation for the acceleration due to self-gravity is
\begin{equation}\label{poisson0}
-\frac{\partial\psi}{\partial r}(r)\approx 2G\int\limits_{-\infty}^{\infty}
\frac{\Sigmap(r+\Delta r)\Delta r}
{(\Delta r)^2 +({H_{\rm p,0}})^2}~d(\Delta r),
\end{equation}
which is equivalent to its radial fourier transform,
\begin{equation}\label{poisson}
\tilde\psi(k)= \int\limits_{\infty}^\infty
\psi(r+\Delta r)\,e^{-ik\Delta r}\,
d(\Delta r)= -\frac{2\pi G}{|k|}e^{-|k|H_{\rm p0}}\,\tilde\Sigma(k),
\end{equation}
if the main contribution comes from $|\Delta r|\ll r$.
The terms involving ${H_{\rm p,0}}$ soften the
force on scales smaller than the equilibrium dust-layer thickness.
Perhaps one ought to use the actual local thickness $\hp(r,t)$ rather
than its equilibrium value, but then the force could not be
calculated so easily by fourier transforms.  It seems pointless
to accept this complication since the exact force softening depends
upon the detailed shape of the vertical density profile, which is beyond
the scope of the height-integrated treatment.

The height-integrated pressure involves the weight of the overlying dust:
\begin{equation}\label{pressure}
\Pi(r,t)= \int\limits_\infty^\infty dz
\int\limits_z^\infty dz'\,\rhop(r,t,z')\Omega^2 z' = \Sigmap\hp^2\Omega^2,
\end{equation}
neglecting a factor of order unity
that also depends upon the vertical density profile.
This formula presupposes vertical hydrostatic equilibrium, which
is justified by the thinness of the dust layer and the slowness of
its motions.  Only the dust density contributes
to the excess pressure since the gas is neutrally buoyant.

We assume that the gas \emph{within} the dust layer is trapped there
on timescales shorter than the settling time (\ref{gamma_s}).
The internal (thermal) energy of this gas
is large compared to the kinetic energy in the radial motions of interest,
even if the trapped gas is a minor contributor to the volume mass density.
On short timescales, therefore, the dust layer is 
is three-dimensionally incompressible:
\begin{equation}\label{Pieqn}
\frac{\delta\Sigmap}{\Sigmap}\approx\frac{\delta\hp}{\hp}~~~\mbox{and}~~~
\frac{\delta\Pi}{\Pi}\approx 3\frac{\delta\Sigmap}{\Sigmap}.
\end{equation}
But $\hp$ should relax towards equilibrium on the settling time
$\gamma_{\rm s}^{-1}$ [eq.~(\ref{gamma_s})], restoring $\Pi$ to its
equilibrium value.
These considerations suggest the following dynamical equation for $\Pi$:
\begin{equation}
\left(\frac{\partial}{\partial t}+U\frac{\partial}{\partial r}\right)\Pi
= -3\Pi\frac{\partial U}{\partial r} - 2\gamma_{\rm s}(\Pi-\Pi_0),
\end{equation}
where ${\Pi_0}$ is the equilibrium pressure.

For the purpose of nonlinear simulations, it is useful to re-cast
the dynamical equations (\ref{eulerV}), (\ref{eulerU}), and (\ref{Pieqn})
in ``conservative'' form.  Multiplying eq.~(\ref{eulerV}) by $\Sigmap$
and eq.~(\ref{masseqn}) by $V$ and adding the results, we have
\begin{displaymath}
\frac{\partial(\Sigmap V)}{\partial t} +
\frac{\partial(U\Sigmap V)}{\partial r} -
V\nu\frac{\partial^2\Sigmap}{\partial r^2}
= -\frac{\Omega\Sigmap U}{2} 
-\,\frac{\rhog \Delta v}{\Rey}\,(V+\eta\vk).
\end{displaymath}
The form of the term involving $\nu$ results from our assumption
that mass diffuses radially but velocities (momentum per unit mass) do not.
This is unreasonable, since one already assumes that \emph{vertical}
momentum diffusion is driven primarily by vertical shear.
It is very unclear, however, exactly what combination of radial
derivatives of $V$ and $\Sigmap$ to use above.
There is a controversy whether hydrodynamic turbulence in a differentially
rotating disk attempts to equalize angular velocity, $(\vk+V)/r$,
or specific angular momentum, $r(\vk+V)$ \citep{BH98}.
Without taking a stand on this issue, we replace the diffusive term above
with
\begin{displaymath}
-\,\nu\frac{\partial(\Sigmap V)}{\partial r}
\end{displaymath}
for the form of the diffusive term, and corresponding forms for
the other dynamical variables.   This leads to somewhat simpler
equations than other plausible choices.  More importantly,
the radial diffusion is then unambiguously stabilizing
at short wavelengths; this is a nontrivial result, because
some other plausible forms of the effective viscosity actually \emph{cause}
axisymmetric instability \citep{ST95}.

Our final set of dynamical equations is therefore
\begin{eqnarray}
\frac{\partial(\Sigmap V)}{\partial t} &+&
\frac{\partial}{\partial r}\left(U\Sigmap V -\nu\frac{\partial(\Sigmap V)}
{\partial r}\right)\nonumber\\
\qquad &=&  -\frac{\Omega\Sigmap U}{2} 
-\,\frac{\rhog \Delta v}{\Rey}\,(V+\eta\vk),
\label{consV}\\
\frac{\partial(\Sigmap U)}{\partial t} &+&
\frac{\partial}{\partial r}\left(\Sigmap U^2 -\nu\frac{\partial(\Sigmap U)}
{\partial r}\right)\nonumber\\
\qquad &=& 
2\Omega\Sigmap V -\frac{\rhog|\Delta v|}{\Rey} U
 -\Sigmap\frac{\partial\psi}{\partial r}
-\frac{\partial\Pi}{\partial r},
\label{consU}\\
\frac{\partial\Sigmap}{\partial t} &+& \frac{\partial}{\partial r}
\left(U\Sigmap - \nu\frac{\partial\Sigmap}
{\partial r}\right) = 0,
\label{consM}\\
\frac{\partial\Pi}{\partial t} &+&
\frac{\partial}{\partial r}\left(U\Pi -\nu\frac{\partial\Pi}
{\partial r}\right) =
 -2\Pi\frac{\partial U}{\partial r} - 2\gamma_{\rm s}(\Pi-\Pi_0),
\label{consP}
\end{eqnarray}
which must be supplemented by Poisson's equation (\ref{poisson}).

\subsection{Dimensionless units}

It is convenient to adopt the system of units and dimensionless
variables described in Table~II, in which $r_0$ is a radius
characteristic of the region of interest.
\marginpar{\small\it Table II goes here}
The numerical values in the final column of the Table
were computed for  a minimum-mass solar nebula in ``equilibrium''
with $r_0=1\au$, $r_{\rm p}=10\cm$, $\Rey=180$, and $\Sc=3$.
In these dimensionless units, equations (\ref{consV})-(\ref{consP})
and (\ref{poisson}) become
\begin{eqnarray}
\label{dimV}
\frac{\partial (\Sigma_*V_*)}{\partial t_*} &+& \frac{\partial}{\partial x_*}
\left(\Sigma_*U_*V_* -\nu_*\frac{\pd(\Sigma_*V_*)}{\pd x_*}\right) =
 -\frac{1}{2}\Sigma_*U_*  -(1+V_*)\frac{R_*}{2},\\
\label{dimU}
\frac{\partial(\Sigma_*U_*)}{\partial t_*} &+& \frac{\partial}{\partial x_*}
\left(\Sigma_*U_*^2 -\nu_*\frac{\partial(\Sigma_*U_*)}{\partial x_*}\right)
= 2\Sigma_*V_* - U_*\frac{R_*}{2} -\Sigma_*\frac{\partial\psi_*}{\partial x_*}
-\frac{\partial\Pi_*}{\partial x_*},\nonumber\\ \\
\label{dimS}
\frac{\partial\Sigma_*}{\partial t_*} &+& \frac{\partial}{\partial x_*}
\left(\Sigma_*U_*
-\nu_*\frac{\partial \Sigma_*}{\partial x_*}\right) = 0,\\
\label{dimPi}
\frac{\partial\Pi_*}{\partial t} &+&
\frac{\partial}{\partial x_*}\left(U_*\Pi_* -\nu_*\frac{\partial\Pi_*}
{\partial x_*}\right) =
 -2\Pi_*\frac{\partial U_*}{\partial x_*} - 2\gamma_*(\Pi_*-\Pi_{*0}),\\
\label{dimpsi}
\tilde\psi(k_*,t_*)&=& -\frac{2}{q_*|k_*|}e^{-|k_*|H_{*0}}
\tilde\Sigma(k_*,t_*).
\end{eqnarray}
We have used the abbreviation
\begin{equation}\label{Rdef}
R_*\equiv\sqrt{U_*^2+(1+V_*)^2}
\end{equation}
for the dimensionless form of the relative velocity (\ref{Delta_v}).

\subsection{Constant states}

We will want to perform a linear stability analysis of 
eqs.~(\ref{dimV})-(\ref{dimpsi}) for small departures from
a background situation in which the dependent variables are
constant with respect to both $x_*$ and $t_*$.  We call this a
``constant state.''
When all of the derivative terms are set to zero, we have
$\Pi_*\to\Pi_{*0}$, $\psi_*\to$ an irrelevant constant,
and two algebraic relations,
\begin{displaymath}
 \frac{1}{2}\Sigma_*U_*  +(1+V_*)\frac{R_*}{2} ~=~0~=~
2\Sigma_*V_* - U_*\frac{R_*}{2}.
\end{displaymath}
This leaves a one-parameter family of possible choices for
$(V_*,U_*,\Sigma_*)$.  Astrophysically, the choice is determined
by the dust-to-gas ratio; at solar abundance, this is $\approx 1/200$,
leading to $\Sigma_{*0}\approx 60$ at $1~\au$ as shown in Table~II.
While the global ratio is fixed by the initial conditions,
the local one may vary, since even in the constant state, the
dust drifts radially inward with respect to the gas.
Depending upon the radial profile of gas density and temperature
(which we have approximated by constants in our local approximation),
this may lead to an enrichment or dilution of the dust at small radii.

So it is interesting to consider, at least briefly, the full range
of possible constant states.  The simplest equations result if one
expresses $\Sigma_*$ and $U_*$ in terms of $V_*$.
The physically allowed range is
\begin{equation}\label{Vlimits}
-1< V_* < 0,
\end{equation}
and
\begin{eqnarray}\label{constates}
U_* &=& -\sqrt{-4V_*(1+V_*)},\nonumber\\
\Sigma_* &=& (1+V_*)\sqrt{\frac{1-3V_*}{-4V_*}}.
\end{eqnarray}
The dimensionless surface density is a monotonic function of $V_*$,
but the radial velocity $U_*$ is most negative at $V_*=-1/2$,
where $-U_*=1=\Sigma_*$.
(In physical units, this would imply an orbital decay time
$r_0/|U|$ of only $160~\yr$ at $1~\au$!)
In the limit $V_*\to-1$ and $\Sigma_*\to\infty$, drag is negligible
because of the large inertia of the dust, which then follows
exact keplerian orbits.  In the opposite limit $V_*\to 0$ and
$\Sigma_*\to 0$, the inertia of the dust is negligible and drag locks
it to the gas.  In both cases, the radial drift velocity vanishes.
The simplicity of equations (\ref{constates}) motivated the choice of
units shown in Table~II, even though the nominal astrophysical
values of the dimensionless variables are far from unity in this system.

A very important quantity is the radial mass flux,
\begin{equation}\label{fluxdef}
F_*\equiv \Sigma_*U_* = -\sqrt{(1+V_*)^3(1-3V_*)} \approx -1 + O(V_*^2)
~~~\mbox{if}~|V_*|\ll 1,
\end{equation}
so the mass flux is almost independent of the dust surface density in
the astrophysically relevant parameter regime.
Mass is the only globally conserved quantity in the dust layer---angular
momentum is ``lost'' to the gas---and
we may imagine that $F_*$ is fixed by a boundary condition at large
radius.  The instabilities found below may be related to 
the fact that such a boundary condition constrains the constant state
only very weakly.

\section{LINEAR STABILITY}
\label{sec:linear}

\S \ref{sec:linear}.1 analyzes 
a simplified problem in which
both the self-gravity and the pressure of the dust layer are ignored.
These simplifications allow the dispersion relation to be presented
relatively easily in closed form.
\S \ref{sec:linear}.2 presents numerical growth rates that incorporate
both of these effects.

\subsection{Analytic treatment without selfgravity or pressure}

The simplified analysis is interesting for several reasons.  First, it
shows that self-gravity is not required for instability, the primary
cause of which is the competition between drag and inertial forces.
Second, the analysis provides a check on our numerically computed
growth rates.  Finally, there are parameter regimes in which the
neglected effects are unimportant---\emph{e.g.}, pressure is
inconsequential for large particles, which settle quickly to the
midplane.

With $\Pi_*$ and $\psi_*$ neglected, only three of equations 
(\ref{dimV})-(\ref{dimpsi}) are needed.  We linearize the remaining
three about an arbitrary constant state (\ref{constates})
and assume that the first-order quantities have the dependence
\begin{equation}\label{delconst}
(\Delta U_*,\Delta V_*,\Delta \Sigma_*) \propto \exp\left[
(z-\nu_* k_*^2)t_* + ik_*x_*\right].
\end{equation}
After considerable algebra (aided by commercial
symbolic-manipulation software)
we obtain a cubic dispersion relation for the shifted complex
growth rate $z$:
\begin{eqnarray}\label{condisp}
0&=&\left[(1+4\,w^{2})(1+w^{2})^{3}\right]{z}^{3} +\left[3w
(w^{2}+1 -2ik_*)(1+4w^{2})(1+w^{2})^{2}\right]{z}^{2}\nonumber\\
&-&\left[ (48{k_*}^{2}w^{4}+12{k_*}^{2}w^{2}-8w^{8}-19w^{6}-15w^{4}
-5w^{2}-1+56{i}w^{6}k_* \right.\nonumber\\
&~&\left. \qquad +70{i}w^{4}k_*+14{i}w^{2}k_*)(1+w^{2})\right]z \\
&+& 8ik_*w^3\left (-6w^{4}+
8{i}w^{4}k_*+2{i}k_*+4{k_*}^{2}w^{2}-3w^{2}+{k_*}^
{2}-3w^{6}+10{i}w^{2}k_*\right ).\nonumber
\end{eqnarray}
To avoid radicals in the coefficients, we have
introduced an auxiliary variable $w$ defined by
\begin{equation}\label{wdef}
V_* = -\frac{w^2}{1+w^2},~~~0<w<\infty.
\end{equation}
Nevertheless, the dispersion relation is cumbersome and best understood
in limiting cases.

First, as $w\to\infty$, \emph{i.e.} the state $(U_*,V_*,\Sigma_*)\to(0,-1,0)$
where the dust layer is pinned to the gas, there are two rapidly
damped roots $z\approx -w$ and $z\approx -2w$.
The interesting root is the third one:
\begin{equation}\label{fastroot}
z\approx 3ik_* w^{-1} +\frac{1}{2}k_*^2 w^{-3},~~~(w\to\infty),
\end{equation}
where we show only the leading order terms in the real and imaginary parts.
The actual growth rate, however,
is $\mbox{Real}(z)-\nu_* k_*^2$.  Therefore, since $\Sigma_*\approx w^{-2}$
in this limit, the system is stable if $\Sigma_*<(2\nu_*)^{2/3}$.

The more important limit is $w\to0$, $(U_*,V_*,\Sigma_*)\to(0,0,\infty)$.
If one assumes that $k_*$ and $z$ are of order unity and sets $w=0$ above,
then the dispersion relation (\ref{condisp}) reduces to $z^3 +z=0$.
This has two purely imaginary roots $z=\pm i$, corresponding to modes
that will damp for any $\nu_*>0$.  To study the root near $z\approx 0$,
one has to go to higher order:
\begin{equation}\label{slowroot}
z\approx (2k_*^2-ik_*^3)\Sigma_*^{-3}
~~~[w\to0,~\Sigma_*\to\infty,~k_*\sim O(1)].
\end{equation}
This indicates instability if $\Sigma_*<(2/\nu_*)^{1/3}$.
In extracting the approximate root (\ref{slowroot}) from 
eq.~(\ref{condisp}), we assumed that $k_*$ is of order unity.
But if we take $k_*= q/w$ with $q\sim O(1)$ as $w\to0$, a different
ordering of terms results, and the dispersion relation reduces
approximately to
\begin{equation}\label{randomcubic}
(\omega')^3 -(\omega') + 2q=0,~~~\omega'\equiv iz +2q.
\end{equation}
Clearly, if $|q|$ is sufficiently large, then there are a pair
of complex-conjugate roots for $\omega'$, implying the existence
of a root for $z$ with positive real part.
The corresponding growth rate is approximately
\begin{displaymath}
3^{1/2}2^{-2/3}q^{1/3}-\nu_* (q/w)^2.
\end{displaymath}
The maximum growth rate is therefore
\begin{equation}\label{fastroot2}
\mbox{Real}(z)_{\rm max}\approx 0.49\Sigma_*^{-2/5}\nu_*^{-1/5}
~~~~(\Sigma_*\to\infty).
\end{equation}
which is achieved at $q\approx 0.31(\Sigma_*^2\nu_*)^{-3/5}$.
But (\ref{randomcubic}) has complex roots only if $|q|>3^{-3/2}$,
and so this growing mode exists only if $\Sigma_*<1.5\nu_*^{-1/2}$,
approximately.

Scrutiny of Table~II shows that the combination $\Sigma_*^2\nu_*$
is independent of the critical Reynolds number $\Rey$.
Therefore, the maximum growth rate is formally independent of this
most uncertain parameter.
The same is true for the ratio $2\pi(k_*H_{*0})^{-1}$ of wavelength to
dust-layer thickness at maximum growth.

To summarize,
if $0<\nu_*\ll 1$, then constant states are unstable when
$\nu_*^{2/3}\lesssim\Sigma_*\lesssim\nu_*^{-1/2}$.
Selfgravity will probably enlarge the range of instability, while
pressure will probably reduce it, but both effects introduce
additional uncertain parameters.

\subsection{Numerical Growth Rates}

As a complement to the analytic treatment of the previous section, we
performed a numerical linear stability analysis of the governing
equations so as to investigate the effects of pressure and
self-gravity upon the expected growth rates. We began our analysis by
transforming the governing equations into Fourier space. In this way,
the gravitational potential can be eliminated in favor of the surface
density via Poisson's eq.~(\ref{poisson}). We next recast the equations by
introducing new variables for the radial and azimuthal mass flux; $f
\equiv \Sigma _* U_* : j \equiv \Sigma _* V_*$. We then proceeded to
linearize the governing equations about the constant states by adding
a perturbation to each dynamic variable (ie. $f \rightarrow f + \Delta
f$), expanding, and then only retaining those terms which are linear
in the perturbations. Thereby, with $\hat{z} \equiv \pd/\pd t_* - \nu
_* \pd^2/\pd x_*^2$, we wrote the system of linearized governing
equations in the form
\begin{displaymath}
\hat{z}(\vy) = \lin \vy,
\end{displaymath}
where $ \vy = (\Delta f,\Delta j,\Delta \Sigma_*,\Delta \Pi_*)$ is the
vector of first-order dynamic variables. Then, given the form of the
($x_*,t_*$) dependence (\ref{delconst}), it
follows that the eigenvalues of the linear operator matrix,
$\lambda_{\lin}$, correspond to dimensionless growth rates =
Re($\lambda_{\lin}$) - $\nu _* k_*^2$.

This analysis was performed for a range of plausible nebula
parameters, specifically; $r$ (orbital radius) $\in$ \{0.1 AU, 1.0 AU,
10 AU\}, $r_p$ (particle size) $\in$ \{0 cm, 10 cm, 100 cm\},
$\mathrm{Re}_* \in$ \{45,70,180,500\}. Figure 1 shows the growth rate
as a function of wavenumber for various parameter values.
For reference,
the unperturbed surface density in our model is about an order of
magnitude lower than the critical surface density required for the
Goldreich-Ward dynamical instability.
\marginpar{\small\it Fig.~1}

For most choices of the background parameters, 
the growth rate is largest at the shortest available wavelengths.
We present results down to a wavelength comparable
to the dust layer thickness, as the validity of the height-integrated 
approach is questionable beyond this
point. There are a few sets of nebula parameters, however,
for which the peak growth rate occurs on a significantly
larger spatial scale. Fig. 2 shows the growth rates for nebula
parameters $r = 0.1$ AU, $r_p = 10$ cm, Re$_* = 180 $. Again, rates
are shown with and without self-gravity. These growth rates are
notably different from those at other parameter values in that they
are considerably lower and become negative for all wavenumbers in the
absence of self-gravity.

It is our interpretation that there are two
distinct secular instabilities in this system. One is the
self-gravitating secular instability discussed in \S 1.
It has a relatively low growth rate and a long radial wavelength
compared to the dust-layer thickness.
This is the instability which appears in Fig. 2.
\marginpar{\small\it Fig. 2}
The other is the drag-induced instability discussed in the previous
section, which has larger growth rates and shorter wavelengths, and
is exhibited in Fig. 1.
Probably because of its large growth rate, the drag instability is
found over a wider range of relevant nebular and particle parameters.
The rest of the results in this section refer to instances in
which the drag instability dominates.

Table III. presents the peak growth rate achieved for a number of other
prominent sets of nebula parameters. Generally, the
dimensionless growth rates
\marginpar{\small\it Table III here}

i) increase monotonically with increasing orbital radius. An
increase in orbital radius implies a decrease in the surface dust
density and an increase in the dust scale height. [The orbital radius
also effects the stopping time by determining the drag regime.]

ii) increase with increasing particle size. An increase in particle
size implies an increase in the settling rate.

iii) increase with increasing critical Reynolds number. An increase
in Reynolds number implies an increase in the dust surface density and
a decrease in the dust scale height.

We repeated the above analysis with the terms relevant to pressure and
self-gravity selectively neglected.\footnote{Recall that in this context,
``pressure'' means the excess height-integrated pressure due to the
weight of the local dust column, cf. \S\ref{sec:eqns}.3.}
As expected, the growth rates are larger without
pressure and smaller without self-gravity.
But the changes are typically on the order of ten to twenty percent.
When both pressure and self-gravity are neglected,
we recovere the growth rates predicted by the cubic dispersion relation
(\ref{condisp}).

In summary, our linear analysis reveals a drag-induced instability
which grows most rapidly on short wavelengths at large orbital radii,
for large particles, and high critical Reynolds number.

\section{DISCUSSION}
\label{sec:discuss}

We have shown that for plausible nebular parameters,
the growth rate of the drag instability can be comparable to
the orbital frequency $\Omega$.

The instability is rather robust.  The following toy model shows that
the \emph{tendency} to instability---as opposed to the actual growth
rate---does not depend upon the more intricate and uncertain
assumptions that we have made in formulating our dynamical equations.
Consider a one-dimensional distribution of mass $\sigma(x,t)$ per unit
length that is subject to a fixed ``gravitational'' acceleration $g$
and an opposing ``frictional'' force per unit length $-f(\sigma)v$.  Here
$\sigma g$ is a proxy for the gravitational and coriolis terms in the
dust-layer equation of motion---forces that are proportional to
mass---whereas $f$ stands for the drag force on the surface of the
dust layer, which is essentially independent of the surface mass
density.  The equations for mass and velocity ($v$) are
\begin{eqnarray*}
\pd_t\sigma + \pd_x(\sigma v) &=& 0,\\
\pd_t v + v\pd_x v &=& g ~-v\,f(\sigma)/\sigma,
\end{eqnarray*}
This system has equilibria with uniform velocity $v_0$ and
surface density $\sigma_0$ satisfying $\sigma_0 g= f(\sigma_0)$.
Without loss of generality, $v_0=0$.  Linear perturbations
with $(x,t)$ dependence $\exp(st+ikx)$ have the dispersion relation
\begin{displaymath}
s^2 = -ik\frac{d}{d\ln\sigma}\left.\left(\frac{f}{\sigma}\right)
\right|_{\sigma=\sigma_0}
\end{displaymath}
Clearly, there will always be an unstable root, $\mbox{Real}(s)>0$, 
regardless of the details of the friction function $f(\sigma)$, unless
the frictional force is strictly proportional to mass.
One could certainly complicate to this model to stabilize
it, but the point is that a balance between inertial
and non-inertial forces tends to be unstable.

The one assumption that is truly critical for instability
is that drag should act coherently  upon the dust layer as whole.
If the drag acts upon each particle independently, 
collective instability cannot result from drag alone.
The assumption of collective drag, introduced by GW,
is reasonable if the gas within the dust layer is substantially entrained
by the dust, which requires that the turbulent wakes of particles
overlap before they mix with the gas above and below the dust layer.
The net drag will then be less than it would be if each particle
independently encountered the full headwind from the subkeplerian gas.
Thus, a necessary condition
for the drag to be treated as a collective effect is that the
force per unit area (\ref{Fdrag}) be less than $N\times f(\rp)$,
where $f(\rp)$ is the drag on an isolated (spherical) particle
of radius $\rp$, and $N$ is the number of such particles per unit
area.
For simplicity, let the dust layer be composed entirely of particles
having this radius; then the surface density of the layer is related
to $\rp$ by
\begin{equation}\label{Ndef}
\frac{4\pi}{3}\rhos\rp^3\,N= \Sigmap,
\end{equation}
where $\rhos\approx 1~\mbox{g cm}^{-3}$ is the density of the solid
material.
In the size regime $\rp\sim 10-100~\cm$, the drag on this particle
can be estimated from Stokes formula,
\begin{equation}\label{fStokes}
f(\rp) = 6\pi\mu_{\rm g}\rp\Delta v,
\end{equation}
with dynamical viscosity $\mu_{\rm g}\approx 0.9\times 10^{-4}~\mbox{g/cm-s}$
at $300$~K.
Thus
\begin{equation}\label{Fratio}
\frac{F_{\rm drag}^{\rm collective}}{N f}\approx
\frac{2\rp^2\rhos\rhog\Delta V}{9\Rey\Sigmap\mu_{\rm g}}
\approx 0.08 \left(\frac{\Rey}{180}\right)^{-1}
\left(\frac{a}{1\,\au}\right)^{-1}
\left(\frac{\rp}{100\,\cm}\right)^2,
\end{equation}
using the nebular parameters and scalings of equations (\ref{MMSN})
and Table I, and assuming that $\Delta V\approx c^2/\vk$.
Actually, the force ratio (\ref{Fratio}) is independent
of the temperature and density profile of nebula as long as the
relative abundance $\Sigmag/\Sigmap$ of gas and dust is constant.
We conclude from this that a collective treatment of the drag is
appropriate as long as $\rp\lesssim 3~\mbox{m}$.

Several groups have previously analyzed the stability of disks when
gas and dust are treated as dynamically distinct components, but
none of them seem to have identified the drag instability that we
have emphasized, and it is interesting to ask why not.
\cite{CFM} performed a stability analysis local to the interior
of the dust layer, where the dust was treated as uniformly mixed into
the gas but enhanced in abundance by sedimentation.  Their vertically
local treatement overlooked the exchange of angular momentum between
the layer as a whole and the relatively dust-free overlying gas.  The
approach of \cite{SN88} was similar, except that they allowed for more
general perturbed velocities within the layer (not necessarily
parallel to the midplane).  Both of these groups found instabilities,
but only for dust densities larger than the critical value $\rho_*$
(cf \S 1), so these presumably were modes that rely upon selfgravity.
\cite{NVC} investigated global \emph{nonaxisymmetric} stability
of the disk.  They allowed for a thin dust layer but
represented the drag force by a term strictly proportional to mass, which
suppresses the nonselfgravitating drag instability, 
as the toy model above demonstrates.
\cite{NVC}'s approximations should have captured
the secular selfgravitating mode sketched in \S 1 if they had
considered axisymmetric modes; they may have found a nonaxisymmetric
version of the instability but did not interpret their results along
these lines [see their Fig.~(4b)].

Is the drag instability necessary to form planetesimals, or will
collisional agglomeration may accomplish the task more quickly?
There are at least two issues:
\begin{itemize}
\item (i) Do collisions result in larger particles, \emph{i.e.},
do colliding particles stick or fragment?
\item (ii) How does the collision rate compare with the growth
rate of the drag instability?
\end{itemize}

The first question is difficult because one does not understand
the structure and elastic strength of the particles.  If these
are loose aggregates, fragmentation is likely at relative velocities
$>10~\mbox{m~s}^{-1}$ (Weidenschilling \& Cuzzi 1993, and references
therein).

As for the second question, it is clear that the collision rate is
as fast or probably faster than the drag instabilities.  If all
particles are spheres of radius $\rp$, then the probability
of crossing the dust layer without a collision is
$\sim\exp(-3\Sigmap/\rhop\rp)$, so that the collision rate per
particle
\begin{equation}\label{collrate}
(\Omega t_{\rm coll})^{-1}\sim \frac{\Sigmap}{2\rhop\rp}\approx
0.13\left(\frac{\rp}{\rm 1~m}\right)^{-1}\left(\frac{a}{\au}\right)^{-3/2}
~\yr^{-1}
\end{equation}
for the nebular parameters of Table~I.

Therefore, if most collisions result in sticking, then collisional
agglomeration is probably more important than drag instability.
But, it is interesting that the latter process is at competitive
with collisions.  The collective instability has the attractive
feature that it may take planetesimals directly from sizes
$\sim 10~\cm$ to $\sim 1~\mbox{km}$ in a single step, and thereby
bypass the size regime in which orbital decay (due to aerodynamic
drag on individual planetesimals) is dangerously rapid.

Assuming that the collective instability does operate, it will be
important to understand the nonlinear outcome.  We have performed
time-dependent numerical integrations of the nonlinear system
(\ref{dimV})-(\ref{dimPi}), neglecting selfgravity and pressure.
Typically, the surface density peaks an order of magnitude or more
above the background.  No simple pattern emerges; multiple peaks
develop at slightly different radial drift speeds.  Numerical
stability does not seem to be an issue as long as a suitably short
time step is used, but unfortunately, the detailed nonlinear outcome
depends upon the form chosen for the diffusive terms.
In eqs.~(\ref{dimV})-(\ref{dimPi}), the diffusive terms are in
``flux-conservation'' form, but there is no clear reason to demand
such a form except in the mass equation (\ref{dimPi}), since radial
and azimuthal momentum can be exchanged with the gas.  We found
nonconservative forms that produced the same linear growth rates but
more organized nonlinear evolution.  Because of these uncertainties,
and because selfgravity and nonaxisymmetry would have to be included
in order to understand planetesimal formation,
we shall not give details of our nonlinear experiments.

Nevertheless, we cannot resist using the linear results to speculate
about the mass of the planetesimals that may form by this instability.
Since the linear growth rate appears to peak at a wavelength
of order $4\pi\hp$ (Fig.~1), a plausible estimate for the mass is
\begin{equation}\label{Mplanet}
M_{\rm planetesimal} \sim \Sigmap (2\pi\hp)^2 
\sim 3\times 10^{19}~\mbox{g}.
\end{equation}
The numerical value is evaluated for the surface density and
thickness of the dust layer at $1~\au$ (Table I).
After expulsion of gas and collapse to solid densities,
this corresponds to a planetesimal radius $\sim 10~\mbox{km}$---larger
than for GW's direct gravitational instability
because our dust layer is thicker than theirs.
\section{SUMMARY}\label{sec:sum}

We have found a new collective instability of the dust layer in a
quiescent protostellar disk.  Like the gravitational instability
contemplated by \cite{Safronov} and \cite{GW}, the new one
may collect particles smaller than a meter directly into planetesimals
larger than a kilometer, thereby circumventing a possible difficulty
with rapid orbital decay of the particles.  But the new instability is
driven by drag rather than selfgravity.  In fact, it is driven by the
very shear turbulence that limits the density of the dust layer and
forestalls direct gravitational instability.  We envison a two-stage
mechanism in which drag first concentrates the dust radially into
rings until they become selfgravitating, whereupon the rings fragment
nonaxisymmetrically.
We find that instability is generic when
inertial forces balance frictional forces in a nonrigid mass layer.
Thus, the tendency to instability is independent of the details
of the dynamical equations we have used to describe the dust layer.
To the extent that these equations are correct and that
the parameters of a standard minimum-mass solar nebula are applicable,
the growth time of the drag instability is
of order the orbital period and comparable to the two-body
collision time.  Planetesimals formed by this instability are estimated
to have radii of order 10 km at $1~\au$.

These conclusions are subject to several caveats.
Most importantly, we lack a fundamental understanding of the
exchanges of momentum and mass between the dust and gas layers.
Turbulence is central to these exchanges, and we are unaware of
any directly comparable terrestrial analogs to the dust layer,
which is probably better described as a slurry than the flat
plate envisaged by Goldreich and Ward.
Formally at least, our growth rates are rather insensitive to
the critical Reynolds number $\Rey$, which describes both
the drag on the dust layer and its thickness (\S\ref{sec:eqns}
\& \S\ref{sec:discuss}), and to the particle size.
Most likely, however, these parameters must lie within a limited
range if our dynamical equations are to be valid.
It is vital for the drag instability that the drag on the dust layer
be treated as a collective process.
According to our best estimates, a collective model is appropriate
for particles smaller than about 3 meters at $1~\au$, and this
limiting size depends upon $\Rey$ (\S 4).
On the other hand, when the particles are much smaller than
$10~\cm$, shear-driven turbulence may not be fully developed,
so that the drag and the growth rate may be substantially reduced
(\S\ref{sec:eqns}.2).  Also, the equilibration time of the dust-layer
thickness is a strong function of particle size; there may be
parameter regimes in which drag quickly produces large contrasts in
surface density, but gravitational instability has to wait for
particles to settle and the volume density to increase.
Direct numerical simulations of dust-gas mixtures in three dimensions
may be necessary to answer these questions.

\acknowledgements

This work was supported by the NASA Origins program under grant
NAG5-8385.

%
%
\begin{table}[p]
\begin{center}
{\bf TABLE I}
\end{center}

{\bf Physical parameters of the protoplanetary disk and dust layer: symbols,
meanings, and assumed numerical values.\hfil}\\[1ex]

\renewcommand{\arraystretch}{1.0}
\begin{tabular}{llll}\hline
 Symbol & Meaning & Formula & Nominal Value  ($1\au$)\\
[0.5ex] \hline
~ & ~ & ~& ~\\
$c$ & isothermal sound speed&$\sqrt{k_B T/\bar\mu}$ & $1.0\,
\mbox{km s}^{-1}$ \\
$\mbox{c}_L, c_\delta$ & boundary-layer factors&--- & $1.5,~2.5$\\
$\eta$&azimuthal drift factor & $(\vk-V_{\rm gas})/\vk$ & $10^{-3}$\\
$\hg$ & gas-layer $1\over2$-thickness&$c/\Omega$ &
$5.1\times 10^6\,\mbox{km}$\\
$\hp$ & dust-layer $1\over2$-thickness& $c_L c_\delta\eta r_0/{\rm Re^*}$ 
   & $3.1\times 10^{3}\,\mbox{km}$\\
$\vk$ & keplerian velocity & $\sqrt{GM/r_0}$ & $30\,\mbox{km s}^{-1}$\\
$\Omega$&orbital angular velocity& $\vk/r_0$ & $2\pi\,\mbox{yr}^{-1}$\\
$r$ & orbital radius &---& $1~\mbox{AU}$\\
$\mbox{Re}^*$ & critical Reynolds number &---& $180$\\
$\mbox{Sc}$  & Schmidt number & --- &$1.0$\\
$\Sigmag$ & gas surface density & ---& $1700~\mbox{g cm}^{-2}$ \\
$Z_{\rm p}$ & dust abundance by mass &---& $0.005$ \\
$\Sigmap$ & dust surface density & $Z_{\rm p}\Sigmag$ &$8.5~\mbox{g cm}^{-2}$\\
$U_0$ & radial drift speed & $-2\rhog(\eta \vk)^2/\Rey\Omega\Sigmap$&
 $-1.0\,\mbox{m s}^{-1}$ \\[0.5ex] \hline
\end{tabular}
\end{table}

\begin{table}[p]
\begin{center}
{\bf TABLE II}
\end{center}

{\bf Dimensionless variables and parameters.\hfil}\\[1ex]

\setlength{\tabcolsep}{10pt}
\renewcommand{\arraystretch}{1.0}
\begin{tabular}{llll}\hline
 Symbol & Meaning & Definition & Nominal Value\\ \hline\\[1.0ex]
\multicolumn{4}{c}{\it Dimensional Units}\\ [1.0ex]
$T_1$ & time  &$\Omega^{-1}$ & $1~{\rm yr}/2\pi$\\
$L_1$ & length  & $\eta r_0$& $10^{-3}~\au$\\
$S_1$& surface density&$2\rhog\eta r_0/\Rey$& $0.28~{\rm g~cm}^{-2}$
\\[1.0ex]
\multicolumn{4}{c}{\it Independent variables}\\ [1.0ex]
$t_*$ & time & $t\cdot\Omega$ & --- \\
$x_*$ & radial separation & $(r-r_0)/L_1$ & ---\\
$k_*$ & radial wavenumber & $k\cdot L_1$ & ---\\[1.0ex]
\multicolumn{4}{c}{\it Dependent variables}\\ [1.0ex]
$\Sigma_*$ & surface density & $\Sigmap/S_1$ & $30.$ \\
$U_*$ & radial drift speed & $U/\eta\vk$ & $-0.033$\\
$V_*$ & azimuthal drift &$V/\eta\vk$& $-2.8\times 10^{-4}$\\
$\Pi_*$ & 2D pressure & $\Pi/S_1(\eta\vk)^2$ & $0.013$\\
$\psi_*$ & self-potential & $\psi/(\eta\vk)^2$ & --- \\[1.0ex]
\multicolumn{4}{c}{\it Dimensionless constant parameters }\\ [1.0ex]
$\nu_*$ & diffusivity & $c_L^2/\Rey^2\Sc$ & $2.3\times 10^{-5}$\\
$H_{*0}$ & equil. $\frac{1}{2}$-thickness & $c_\delta c_L/\Rey$ & 
$0.021$\\
$\Pi_{*0}$ & equil. pressure & $\Sigma_{*0}H_{*0}^2$ & $0.013$\\
$q_*$ & self-gravity & $\eta\vk\Omega/\pi GS_1$ & $1.0\times 10^4$\\
$\gamma_*$ & settling rate & $\Omega t_s$ & $0.11$ \\[0.5ex] \hline
\end{tabular}
\end{table}

\begin{table}[p]
\begin{center}
{\bf TABLE III}
\end{center}

{\bf Local axisymmetric instabilities of the dust layer.
Numerical growth rates (column 4) are given in units of local orbital
frequency $\Omega$.\hfil}\\[1ex]

\begin{tabular}{p{1cm}p{1cm}p{1cm}p{2cm}p{2cm}p{5cm}}
$r_p~(\cm)$ & $r~(\au)$ & $\Rey$  & Maximum growth rate
& $e-$folding time (yr) &Comments\\
\hline
& & & & &\\
0 & 0.1 & 180 & 1.8 $\times 10^{-4}$ & 28.0 & \\
& & & & & \\
10 & 0.1 & 45  & 2.1 $\times 10^{-4}$ & 24.0 & long-wavelength instability\\
10 & 0.1 & 180  & 1.9 $\times 10^{-4}$ & 26.5 & \\
10 & 0.1 & 500  & 0.30  & 1.7 $\times 10^{-2}$ &   \\
& & & & & \\ 
100 & 0.1 & 180 & 0.24 & 2.1 $\times 10^{-2}$ & \\
& & & & & \\
0 & 1.0 & 180 & 0.46 & 0.35 & \\
& & & & & \\
10 & 1.0 & 45 & 0.45 & 0.35 & \\
10 & 1.0  & 180 & 0.55 & 0.29 & nominal parameters  \\
10 & 1.0 & 500 & 0.69 & 0.23 & \\
& & & & & \\
100 & 1.0 & 180 & 0.73 & 0.22 &  \\ 
& & & & & \\
0 & 10.0 & 180 & 0.78 & 6.45 & \\
& & & & & \\
10 & 10.0 & 180 & 0.76 & 6.49 & \\
10 & 10.0 & 180 & 0.90 & 5.58 & \\
10 & 10.0 & 180 & 1.01 & 4.96 & \\
& & & & & \\
100 & 10.0 & 180 & 0.94 & 5.31 & \\[0.5ex]\hline
\end{tabular}	
\end{table}
%
%
\begin{figure}[p]
\scalebox{0.9}{\includegraphics[50,144][592,718]{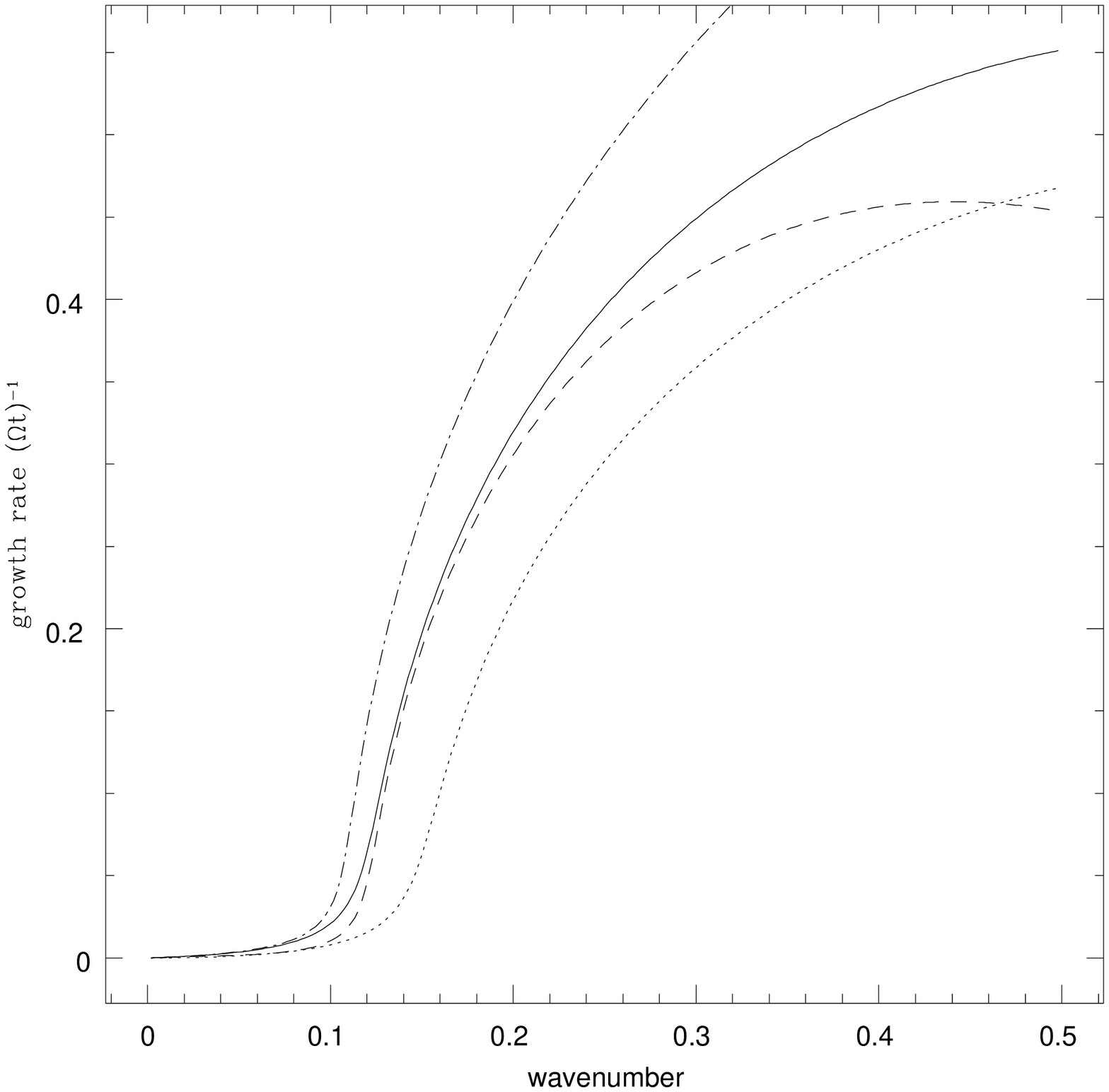}}

\pagestyle{empty} \textbf{FIG. 1.} Dimensionless linear growth rates
for nebula parameters $r = 1$ AU, Re$_* = 180 $. The four curves
represent models for $r_p = 10$ cm with (solid line) and without
(short dashed) self-gravity, as well as for $r_p = 0$ cm(dashed) and
$r_p = 100$ cm (dot-dashed), both with self-gravity.
The units of wavenumber are such that a
wavenumber of 1.0 corresponds to an instability at a wavelength of $2
\pi \hp$.
\end{figure}

\begin{figure}[p]
\scalebox{0.9}{\includegraphics[50,144][592,718]{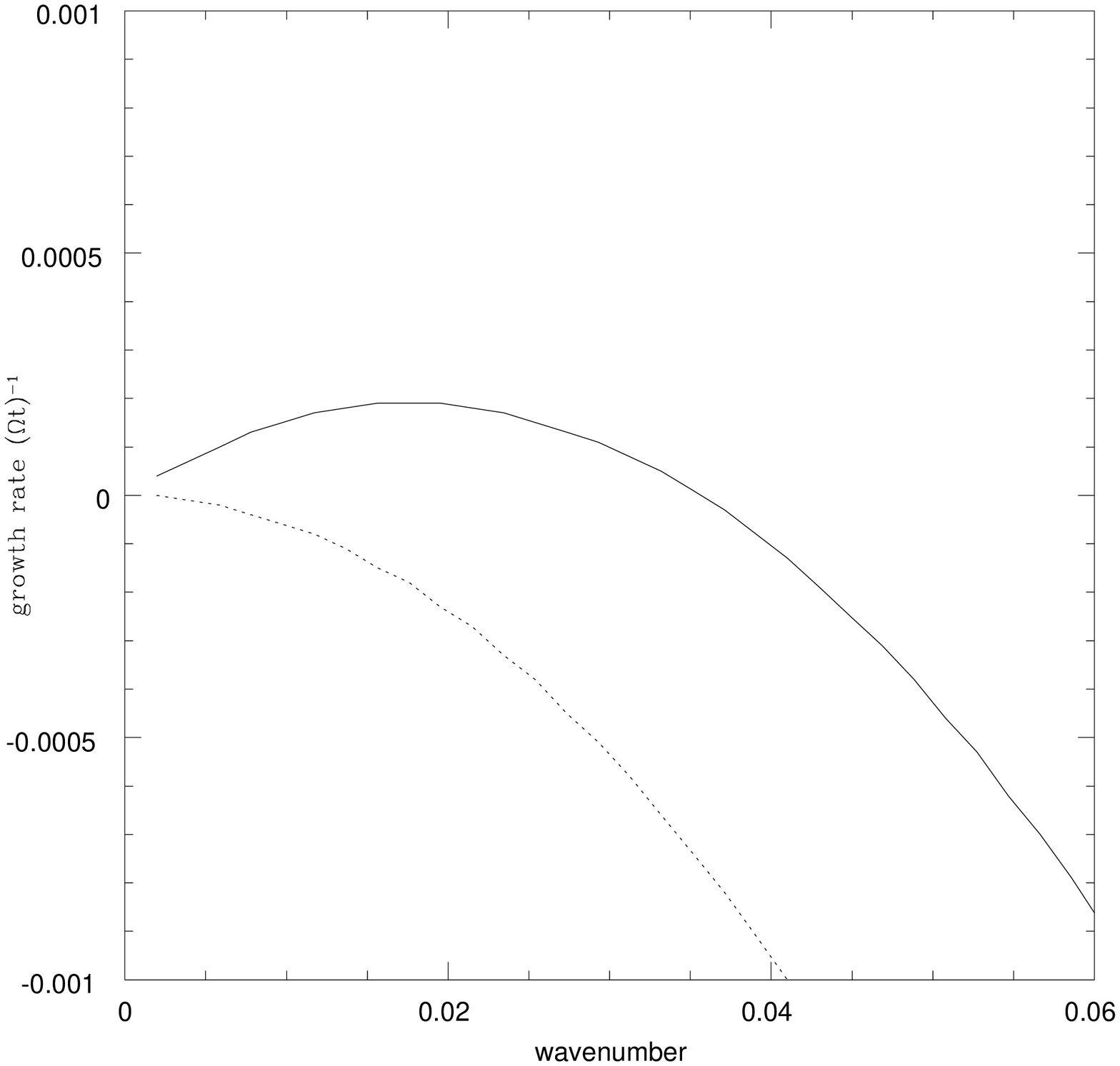}}

\textbf{FIG. 2.} Dimensionless linear growth rates for nebula
parameters $r = 0.1$ AU, $r_p = 10$ cm, Re$_* = 180 $. The two curves
represent models with (solid line) and without (short dashed)
self-gravity. The instability for this set of nebula parameters is
slowly-growing, occurs at long wavelengths, and is not seen in the
absence of self-gravity. The units of wavenumber are such that a
wavenumber of 1.0 corresponds to an instability at a wavelength of $2
\pi H_{*0}$.
\end{figure}

\end{document}